\documentclass[prl,twocolumn,aps,showpacs]{revtex4}
\usepackage{psfig}
\usepackage{bm}

\begin{document}
\date{\today}
\title{Extracting the condensate density from projection experiments with Fermi gases}
\author{A. Perali, P. Pieri, and G.C. Strinati}
\affiliation{Dipartimento di Fisica Universit\`{a} di Camerino, I-62032 
Camerino, Italy}

\begin{abstract}
A debated issue in the physics of the BCS-BEC crossover with trapped Fermi atoms is to identify characteristic properties of the 
superfluid phase.
Recently, a condensate fraction was measured on the 
BCS side of the crossover by sweeping the system 
in a fast (nonadiabatic) way from the BCS to the BEC sides, thus ``projecting'' the initial many-body state onto a molecular condensate.
We analyze here the theoretical implications of these projection experiments, by identifying
the appropriate quantum-mechanical operator associated with the measured quantities and 
relating them to the many-body correlations occurring in the BCS-BEC crossover.
Calculations are presented over wide temperature and coupling ranges, by including pairing fluctuations on top of mean field.
\end{abstract}

\maketitle


The current experimental advances with trapped Fermi atoms have attracted much interest in the physics of the BCS-BEC crossover.
In this context, one of the most debated issues is the unambiguous detection of superfluid properties on the BCS side of the 
crossover.
Several attempts have been made in this direction.
They include absorption images of the ``projected'' density profiles for $^{40}$K \cite{Jin-PRL-2004} and $^{6}$Li \cite{Ketterle-PRL-2004}, 
rf spectroscopy to detect single-particle excitations \cite{Grimm-Science-2004}, and measurements of collective modes 
\cite{Thomas-PRL-2004,Grimm-PRL-2004}.
In addition, a number of schemes to detect superfluid properties on the BCS 
side of the crossover have been proposed, including Josephson 
oscillations~\cite{Josephson} and vortices~\cite{vortices}.

In particular, the experimental procedure of Refs.~\cite{Jin-PRL-2004,Ketterle-PRL-2004} pairwise ``projects'' 
fermionic atoms onto molecules, by preparing the system of trapped Fermi atoms on the BCS side with a tunable 
Fano-Feshbach resonance and then rapidly sweeping the magnetic field to the BEC side. 
In this way, the same two-component fit of density profiles routinely used for Bose gases is exploited to extract from these 
``projected'' density profiles the analog of a condensate fraction, which is now associated with the equilibrium state on the BCS 
side before the sweep took place.

Purpose of this paper is to provide a theoretical interpretation of the experiments of Refs.~\cite{Jin-PRL-2004,Ketterle-PRL-2004}, 
by obtaining the ``projected'' density profiles in terms of the correlation functions of the Fermi gas at equilibrium. 
This will be based on a number of physical assumptions which we associate with the experimental procedure of 
Refs.~\cite{Jin-PRL-2004,Ketterle-PRL-2004}.
Our calculation evidences how the projection procedure amplifies the visibility of the emergence of the condensate as the temperature is lowered below the critical temperature $T_{c}$,
when compared with the ordinary density profiles of Ref.~\cite{PPPS-PRL-04}.
We also attempt an analysis of the ``projected'' density profiles in terms of a two-component fit, in analogy to what is done with 
the experimental data \cite{Jin-PRL-2004,Ketterle-PRL-2004}.
A prediction is further made of a reduced molecular fraction that depends on the initial many-body state, in agreement with a late
experimental evidence \cite{Ketterle-cond-mat-2004}.

Inclusion of pairing fluctuations on top of mean field along the lines of Ref.~\cite{PPS-PRB-04} enables us to cover a wide temperature 
range even in the intermediate- and strong-coupling regimes, in 
contrast to Refs.~\cite{Ho-condensate-2004,Bohn-condensate-2004} where 
only mean field was taken into account.


To account for the ``projected'' density profiles of Refs.~\cite{Jin-PRL-2004,Ketterle-PRL-2004}, we consider the boson-like 
field operator:
\begin{equation}
\Psi_{B}({\bf r}) = \int \! d\bm{ \rho} \, \phi(\rho)  
\psi_{\downarrow}({\bf r}-\bm{ \rho}/2) \psi_{\uparrow}({\bf r}+
\bm{ \rho}/2)\; .
\label{definition-boson-fermion}
\end{equation}
Here, $\psi_{\sigma}(\mathbf{r})$ is a fermion field operator with spin $\sigma$, and $\phi(\mathbf{\rho})$ a real and normalized function
which specifies the probability amplitude for the fermion pair.
An operator of the form (\ref{definition-boson-fermion}) was considered in Ref.~\cite{APS-2003} to obtain the condensate density for 
composite bosons.

Our theoretical analysis of the experiments of Refs.~\cite{Jin-PRL-2004,Ketterle-PRL-2004} is based on the following 
\emph{physical assumptions}, that we infer from the experimental procedure:

\noindent
(i) Atoms of a specific spin state were detected, which originate from the dissociation of molecules after applying an rf pulse. 
    The object of the measurement is thus the bosonic (molecular) density $n_{B}(\mathbf{r})$ at position $\mathbf{r}$  
    (and not the fermionic (atomic) density $n(\mathbf{r})$).
    
\noindent
(ii) A rather large conversion efficiency into molecules results when rapidly sweeping the magnetic field
in the experiments. This suggests that molecules form just past the unitarity limit on the BEC side, where the ``final'' molecular wave function and the many-body correlations for the ``initial''states considered in Refs.~\cite{Jin-PRL-2004,Ketterle-PRL-2004} have maximum 
overlap. Correspondingly, we assume that the wave function in the expression (\ref{definition-boson-fermion}) refers to this ``final'' coupling, and represent it by $\phi_{f}(\mathbf{\rho})$.
     As molecules form in a medium, we take into account the effect of Pauli blocking in analogy with the original Cooper argument \cite{Cooper-1956} and identify $\phi_{f}(\rho)$ with the bound-state solution of the 
     two-body problem, with the condition that its Fourier transform $\phi_{f}(\mathbf{k})$ vanishes when the magnitude of the wave vector 
     $\mathbf{k}$ is smaller than a characteristic value $k_{\mu_{f}}=\sqrt{2 m \mu_{f}}$ for $\mu_{f} > 0$ (
      $m$ being the fermion mass), 
     while no constraint is enforced for $\mu_{f} < 0$.
     Here, the value of the chemical potential $\mu_{f}$ depends on the ``final'' coupling at which the molecular state is 
     assumed to form, thus implying that some sort of local equilibrium can be established around a molecule.
    We shall present our calculations for two ``final'' couplings that bound the interval where the 
maximum overlap occurs, and are at the same time representative of the two cases where $\mu_{f}$ is 
positive or negative.
     
\noindent
(iii) In the experiments, bosonic Thomas-Fermi (TF) profiles for the molecular condensate were extracted from position-dependent density profiles, thus entailing an assumption of thermal equilibrium. 
We assume that this thermal equilibrium corresponds to the state prepared \emph{before} the rapid sweep of the magnetic field.
The validity of this assumption is supported by a recent experimental study of the formation time of a fermion-pair 
condensate\cite{Ketterle-cond-mat-2004}.
In our calculations, we then use $\langle \cdots \rangle_{i}$ as expressions for the thermal averages, where the suffix $i$ stands for ``initial''.
     
All these assumptions are summarized by stating that the \emph{``projected'' bosonic density profile} given by 
\begin{equation}
n_{B}^{fi}(\mathbf{r}) \, = \, \langle \Psi_{B}^{f}(\mathbf{r})^{\dagger} \, \Psi_{B}^{f}(\mathbf{r}) \rangle_{i}     
\label{projected-bosonic-density}       
\end{equation} 
represents the ``in situ'' molecular density which would be measured after the rapid sweep but before the cloud expansion performed in Refs.~\cite{Jin-PRL-2004,Ketterle-PRL-2004} (connection with the density measured after the expansion will be discussed below).
In this expression, the boson-like field operator of Eq.~(\ref{definition-boson-fermion}) contains the \emph{final} molecular wave function 
$\phi_{f}(\mathbf{\rho})$ on the BEC side of the crossover, while the thermal average $\langle \cdots  \rangle_{i}$ is taken with reference 
to the state in which the system was \emph{initially} prepared.

Consistently with our previous work \cite{SPS-2004}, we describe the interaction term of the many-fermion Hamiltonian via an effective 
single-channel model.
The parameter $(k_{F} a_{F})^{-1}$ then drives the crossover from the BCS side (identified by $(k_{F} a_{F})^{-1} \lesssim -1$) 
to the BEC side (identified by $1 \lesssim (k_F a_F)^{-1}$) across the unitarity limit $(k_{F} a_{F})^{-1} = 0$. 
Here, $a_{F}$ is the two-fermion scattering length and the Fermi wave vector $k_{F}$ results by setting $k_{F}^{2}/(2m)$ equal to the noninteracting
Fermi energy. 

The calculation proceeds by expressing the four-fermion field operator in Eq.~(\ref{projected-bosonic-density}) in terms of the 
two-particle Green's function
$\mathcal{G}_{2}(1,2,1',2') \, = \, \langle T_{\tau}[\Psi(1) \, \Psi(2) \, \Psi^{\dagger}(2') \, \Psi^{\dagger}(1')] \rangle$,
where $T_{\tau}$ is the imaginary-time ordering operator.
We have introduced the spinor $\Psi(\mathbf{r}) = (\psi_{\uparrow}(\mathbf{r}),\psi_{\downarrow}^{\dagger}(\mathbf{r}))$ as well as the short-hand notation $1=(\mathbf{r}_{1}, \tau_{1}, \ell_{1})$ with imaginary time $\tau$ and spinor component $\ell$, such that 
$\Psi(1) = \exp\{K \tau_{1}\} \Psi_{\ell_{1}} (\mathbf{r_{1}}) \exp\{- K \tau_{1}\}$.
The thermal average contains the grand-canonical Hamiltonian $K = H - \mu N$ with fermionic chemical potential $\mu$  and is taken in the initial state, as specified above.

The two-particle Green's function $\mathcal{G}_{2}$ is, in turn, expressed in terms of the many-particle T-matrix, by solving formally 
the Bethe-Salpeter equation as follows:
\begin{eqnarray}
& &\mathcal{G}_{2}(1,2,1',2') =\mathcal{G}(1,1') \, \mathcal{G}(2,2') \, - \, \mathcal{G}(1,2') \, \mathcal{G}(2,1')\phantom{1111} \nonumber\\         
& & -  \int \! d3456 \,\, \mathcal{G}(1,3) \, \mathcal{G}(6,1') \, T(3,5;6,4) \, \mathcal{G}(4,2') \, \mathcal{G}(2,5)\phantom{1111} \label{G-T}        
\end{eqnarray}
where $\mathcal{G}(1,1') \, = \, - \, \langle T_{\tau}[\Psi(1) \, \Psi^{\dagger}(1')] \rangle$ is the fermionic single-particle Green's 
function.
Accordingly, the ``projected'' bosonic density (\ref{projected-bosonic-density}) reads:
\begin{equation}
n_{B}^{fi}(\mathbf{r}) \, = \, \int \! d\bm{\rho} \, d\bm{\rho}' \,\, \phi_{f}(\mathbf{\rho}) \,\, \phi_{f}(\mathbf{\rho'}) \,\,
                             \mathcal{G}_{2}^{i}(1,2,1',2')                    \label{pbd-Green-function}       
\end{equation} 
where 
$1  = (\mathbf{r}-\bm{\rho}/2,\tau + 2 \eta,\ell = 2)$,  
$ 2 = (\mathbf{r}+\bm{\rho}'/2,\tau,\ell = 1)$,
$1' = (\mathbf{r}+\bm{\rho}/2,\tau + 3 \eta,\ell = 1)$, 
and $2' = (\mathbf{r}-\bm{\rho}'/2,\tau + \eta,\ell = 2)$ 
($\eta$ being a positive infinitesimal).

Implementation of the above expressions to the trapped case is readily obtained via a local-density approximation, whereby a local gap 
parameter $\Delta(\mathbf{r})$ is introduced and the chemical potential $\mu$ is replaced (whenever it occurs for both ``initial''
and ``final'' couplings) by the quantity $\mu(\mathbf{r}) = \mu - V(\mathbf{r})$ that accounts for the trapping potential 
$V(\mathbf{r})$.


The three terms on the right-hand side of Eq.~(\ref{G-T}) correspond to physically different contributions to the expression 
(\ref{pbd-Green-function}).
In particular, the first term can be written as $|\alpha_{fi}|^{2}$, where 
$\alpha_{fi} = \int \! d\bm{\rho} \, \phi_{f}(\mathbf{\rho}) \mathcal{G}_{12}^{(i)}(\mathbf{\rho},\tau = - \eta)$   
represents the \emph{overlap} between the fermionic correlations (embodied in the anomalous single-particle Green's function 
$\mathcal{G}_{12}^{(i)}$) and the molecular 
wave function $\phi_{f}$.
This contribution vanishes with the gap parameter $\Delta$ when approaching $T_{c}$, and is identified with the \emph{condensate density} 
for composite bosons when both the ``initial'' thermal equilibrium and the ``final'' molecular wave function are taken at the same 
coupling deep in the BEC region \cite{APS-2003}.
Only this contribution was considered in Ref.~\cite{Bohn-condensate-2004} in connection with the experiments of 
Refs.~\cite{Jin-PRL-2004,Ketterle-PRL-2004}.

The second term on the right-hand side of Eq.~(\ref{G-T}) represents fermionic correlations in the normal state, that are relevant in the presence of an underlying Fermi surface.
This term is most sensitive to the ``final'' molecular wave function $\phi_{f}$ being affected by Pauli blocking when $\mu_{f}$ is positive.
This term would be irrelevant if the ``initial'' thermal equilibrium and the ``final'' molecular wave function were
taken deep in the BEC side.
When $\mu_{f}$ is positive, this term can lead to an overestimate of the value of the condensate fraction when the 
``projected'' density profiles are fitted in terms of TF and Gaussian functions, as argued below.
Both this and the previous contribution were considered in Ref.~\cite{Ho-condensate-2004} (where the ``final'' coupling was, however, taken
deep in the BEC region).

The third term on the right-hand side of Eq.~(\ref{G-T}) will be calculated in the following within the off-diagonal BCS-RPA approximation 
considered in Ref.~\cite{APS-2003}.
This contribution is identified with the \emph{noncondensate density} for composite bosons when both the ``initial'' thermal 
equilibrium and the ``final'' molecular wave function are taken at the same coupling deep in the BEC side \cite{APS-2003}.
It is thus of particular importance for increasing temperature when 
approaching the normal phase. 

When $i=f$ deep in the BEC region, the ``projected'' density 
profile $n_{B}^{ii}(\mathbf{r})$ coincides with (half) the ordinary density profile $n(\mathbf{r})$ calculated at the same coupling.
In the following, the values for the local chemical potential and gap parameter, to be inserted in the expression (\ref{pbd-Green-function}) for $n_{B}^{fi}(\mathbf{r})$, are obtained with the theory of 
Ref.~\cite{PPPS-PRL-04} where pairing fluctuations are included on top of mean field.


\begin{figure}
\centerline{\psfig{figure=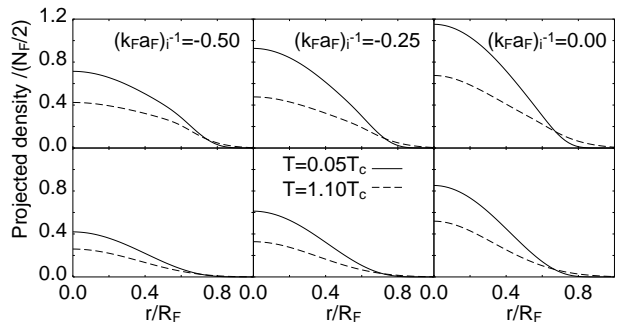,width=7.2cm,angle=-90}}
\caption{Axially-integrated ``projected'' density profiles (in units of $R_F^{-2}$ where $R_F$ is the
fermionic Thomas-Fermi radius) for three 
``initial'' coupling values and for two ``final'' couplings: $0.40$ (upper panel) and $1.50$ (lower panel). 
}
\end{figure}

Figure 1 shows the axially-integrated ``projected'' density profiles calculated for the coupling values 
$(k_{F} a_{F})_{i}^{-1} = (-0.50,-0.25,0.00)$ and for the two 
representative values $0.40$ (upper panel) and $1.50$ (lower panel) of the ``final'' coupling $(k_{F} a_{F})_{f}^{-1}$.
Two characteristic temperatures (just above the critical temperature and near zero temperature) are considered in each case.
Note the marked temperature dependence of the ``projected'' density profiles when entering the superfluid phase, as signaled by the 
emergence of a ``condensate'' component near the center of the trap.
This contrasts the milder dependence (especially on the BCS side) of the density profiles without projection~\cite{PPPS-PRL-04}.
The ``projection'' technique introduced in Ref.~\cite{Jin-PRL-2004} is thus demonstrated to \emph{amplify} the effects due to the
presence of a condensate on the density profiles, which would otherwise be almost temperature independent on the BCS side of 
the crossover.

\begin{figure}
\centerline{\psfig{figure=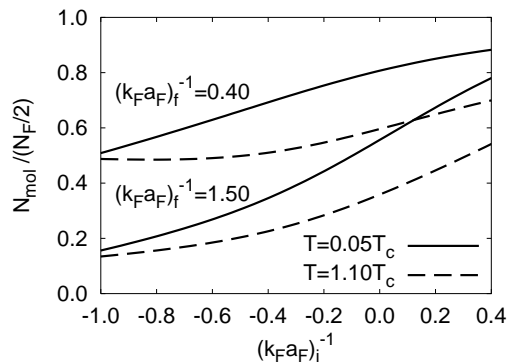,width=7cm,angle=-90}}
\caption{Ratio $N_{mol}/(N_{F}/2)$ vs $(k_{F} a_{F})_i^{-1}$ for two temperatures and ``final'' couplings.}
\end{figure}

In Fig.~1 the densities are normalized to half the total number $N_{F}$ of fermionic atoms.
This number differs, in general, from the total number $N_{mol}$ of molecules obtained by integrating the ``projected'' density profiles.
In particular, $N_{mol}$ can vary significantly when scanning the ``initial'' coupling $(k_{F} a_{F})_i^{-1}$ on the BCS side of the crossover 
for given ``final'' molecular-like state.
This effect is shown in Fig.~2 for the same temperatures and ``final'' 
couplings of Fig.~1.
Our finding that the total number $N_{mol}$ of molecules constitutes only a fraction of the original atom number $N_{F}/2$ for each spin state 
is supported by the experimental results of Refs.~\cite{Jin-PRL-2004} and \cite{Ketterle-PRL-2004}.
In addition, our prediction that the reduced value of the molecular fraction 
depends on the ``initial'' many-body state is in agreement with 
the experimental evidence recently reported in Ref.~\cite{Ketterle-cond-mat-2004}. Note that, for both values of the ``final coupling'', the total number of 
molecules increases upon lowering the temperature. This result indicates that 
the conversion efficiency for the condensate fraction is larger than for
the thermal component, as also observed in 
Ref.~\cite{Ketterle-cond-mat-2004}.

In our procedure, the {\em condensate\/} and {\em noncondensate\/} components of the ``projected'' density profiles are calculated separately.
By our definition, they correspond to the first term and to the remaining terms on the right-hand side of Eq.~(\ref{G-T}), 
respectively.
The \emph{condensate fraction} is obtained accordingly from the ratio of the corresponding areas.
Yet, the total ``projected'' density profiles obtained theoretically could be analyzed in terms of a two-component fit with a TF 
plus a Gaussian function (or, better, a $\mathrm{g}_{3/2}$ function for the Bose gas), in analogy to a standard 
experimental procedure.
This kind of analysis is reported in Fig.~3 for the two low-temperature curves shown on the right panels of Fig.~1.
In both cases, a good overall fit is obtained by the $\chi$-square method.
Separate comparison is also made in the figure with the theoretical condensate and noncondensate components defined above, which 
appears rather 
good for the value $1.50$ of the 
``final'' coupling while for the value $0.40$ an overestimate (of about 
$50 \%$) of the condensate results from the fit.
This discrepancy  stems mostly from the second term on the right-hand side of Eq.~(\ref{G-T}), which contributes to the
TF component of the fit owing to a peculiar shape of the corresponding ``projected'' density profile.

\begin{figure}

\centerline{\psfig{figure=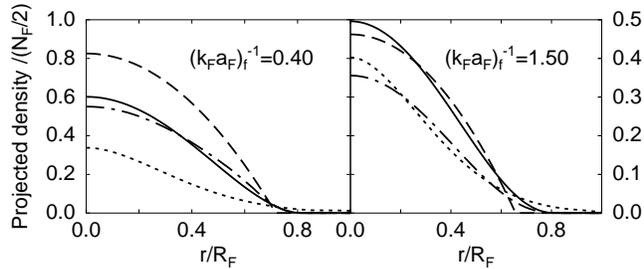,width=8cm,angle=-90}}
\caption{Two-component (TF plus $\mathrm{g}_{3/2}$) fit of the 
axially-integrated ``projected'' 
density profiles (in units of $R_F^{-2}$) at the unitarity limit for $T/T_{c}=0.05$.
The TF (dashed line) and $\mathrm{g}_{3/2}$ (dotted line) components of the 
fit are compared with the theoretical condensate (full line) and 
noncondensate (dashed-dotted line) components.}
\end{figure}

Extention of this analysis to ``initial'' couplings on the BCS side reveals unconventional forms of the 
theoretical condensed and noncondensed contributions to the ``projected'' density profiles, so that the above 
two-component fit fails. 
For negative values of the initial coupling when the two-component fit fails, we have verified that the difference
$N_{mol}(T=0) - N_{mol}(T=1.10 T_c)$ approximately coincides with $N_0(T=0)$ within a relative error not larger than 
$10 \%$ when $(k_F a_F)^{-1}_f=0.4$.
This suggests a practical prescription to extract $N_0(T=0)$ from the values of $N_{mol}$ at low temperature and slightly above the 
critical temperature, without relying on a two-component fit.

In this respect, recall that our theoretical ``projected'' density profiles are calculated when 
molecules just form on the BEC side near the unitarity limit. In the experiments, 
however, a further ramp of the magnetic field is performed together with a subsequent 
cloud expansion. Only at this stage the profile of the cloud is detected. 
Comparison  between theory and experiments is thus meaningful since
the further ramp of the magnetic field and the subsequent cloud expansion are expected 
to have no influence on the values of $N_{mol}$ and 
$N_{0}/N_{mol}$. This is because, by the further ramp, the molecules shrink
following the field adiabatically. They then become tightly bound and weakly 
interacting among themselves, while their local counting is unaffected.
Under these conditions, the condensate fraction, too, should not be modified
by the subsequent expansion as it is the case for an ordinary dilute Bose gas, even though
the expansion may affect the details of the density profiles.

In Fig.~4 the condensate fraction $N_{0}/N_{mol}$, obtained from our theoretical expressions at the lowest temperature $T/T_{c}=0.05$, is plotted vs $(k_{F} a_{F})^{-1}_{i}$ on the BCS side of the crossover.
The data from Refs.~\cite{Jin-PRL-2004} and ~\cite{Ketterle-PRL-2004} are also reported in the figure.
The agreement between the overall trends of the theoretical and experimental curves appears satisfactory, although quantitative discrepancies
result between the two sets of curves.
They might be due to an overestimate of the TF component of the fits in Ref.~\cite{Ketterle-PRL-2004} for the reasons  
discussed in Ref.~\cite{Ketterle-cond-mat-2004}, and to a possible underestimate of the condensate component in Ref.~\cite{Jin-PRL-2004} due to a preferential loss of molecules in the condensate itself.
We have, finally, verified that a linear dependence occurs between $N_{mol}/(N_{F}/2)$ and 
$N_{0}/N_{mol}$ (inset of Fig.~4). A similar linear dependence is 
also reported in Fig.~4(b) of Ref.~\cite{Ketterle-PRL-2004}, although with a 
different definition of $N_{mol}$ that includes also molecular states not 
directly detected in the experiment.


\begin{figure}
\centerline{\psfig{figure=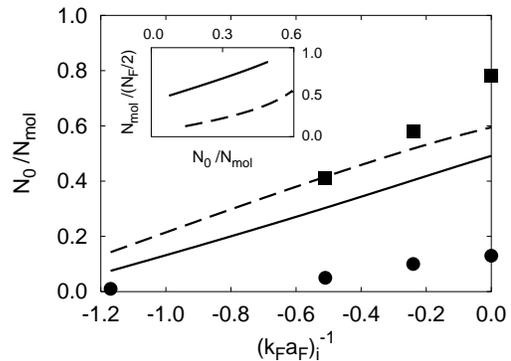,width=7cm,angle=-90}}
\caption{Condensate fraction $N_{0}/N_{mol}$ vs $(k_{F} a_{F})^{-1}_{i}$ on the 
BCS side of the crossover at $T/T_{c}=0.05$ and for the two ``final'' couplings $0.40$ (full line) and $1.50$ (dashed line).
The data from the experiments of Refs.~\cite{Jin-PRL-2004} (dots) and~\cite{Ketterle-PRL-2004} 
(squares) are also reported. The inset 
shows $N_{mol}/(N_{F}/2)$ vs $N_{0}/N_{mol}$ for the same temperature and ``final'' couplings.}
\end{figure}


We are indebted to C. Salomon and H. Stoof for discussions, and to D. Jin
for a critical reading of the manuscript.
This work was partially supported by the Italian MIUR with contract Cofin-2003 ``Complex Systems and Many-Body Problems''.



\end{document}